\documentclass[12pt]{article}
\pdfoutput=1

\usepackage[latin1]{inputenc}
\usepackage[T1]{fontenc}
\usepackage{url}
\usepackage{color}
\usepackage{mathptmx}

\usepackage{floatflt,epsfig}
\usepackage{graphics}

\usepackage{pgf}

\DeclareUrlCommand\email{\urlstyle{rm}}

\pgfdeclareimage[height=6cm,width=6cm]{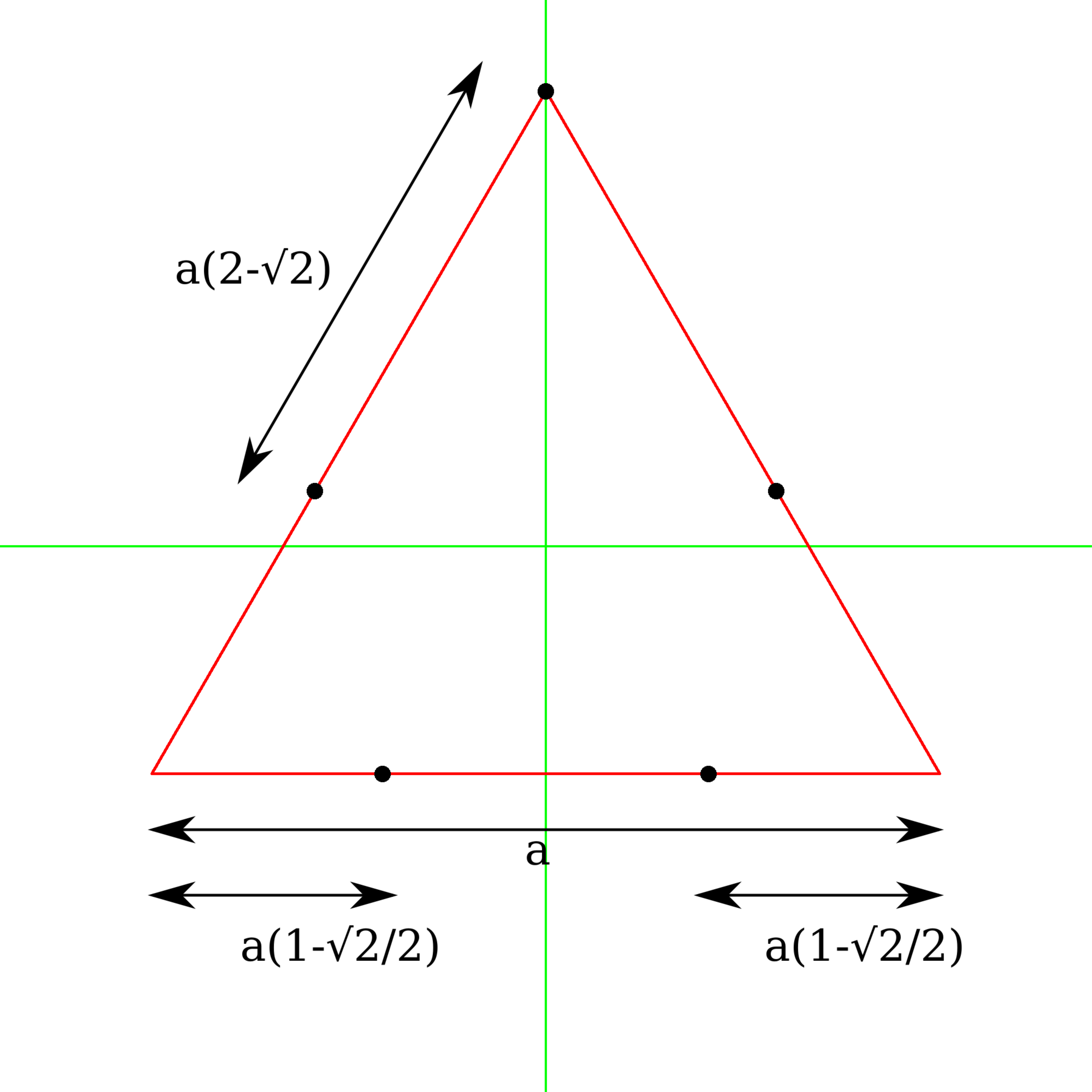}{notreconjecture5points}
\pgfdeclareimage[height=6cm,width=6cm]{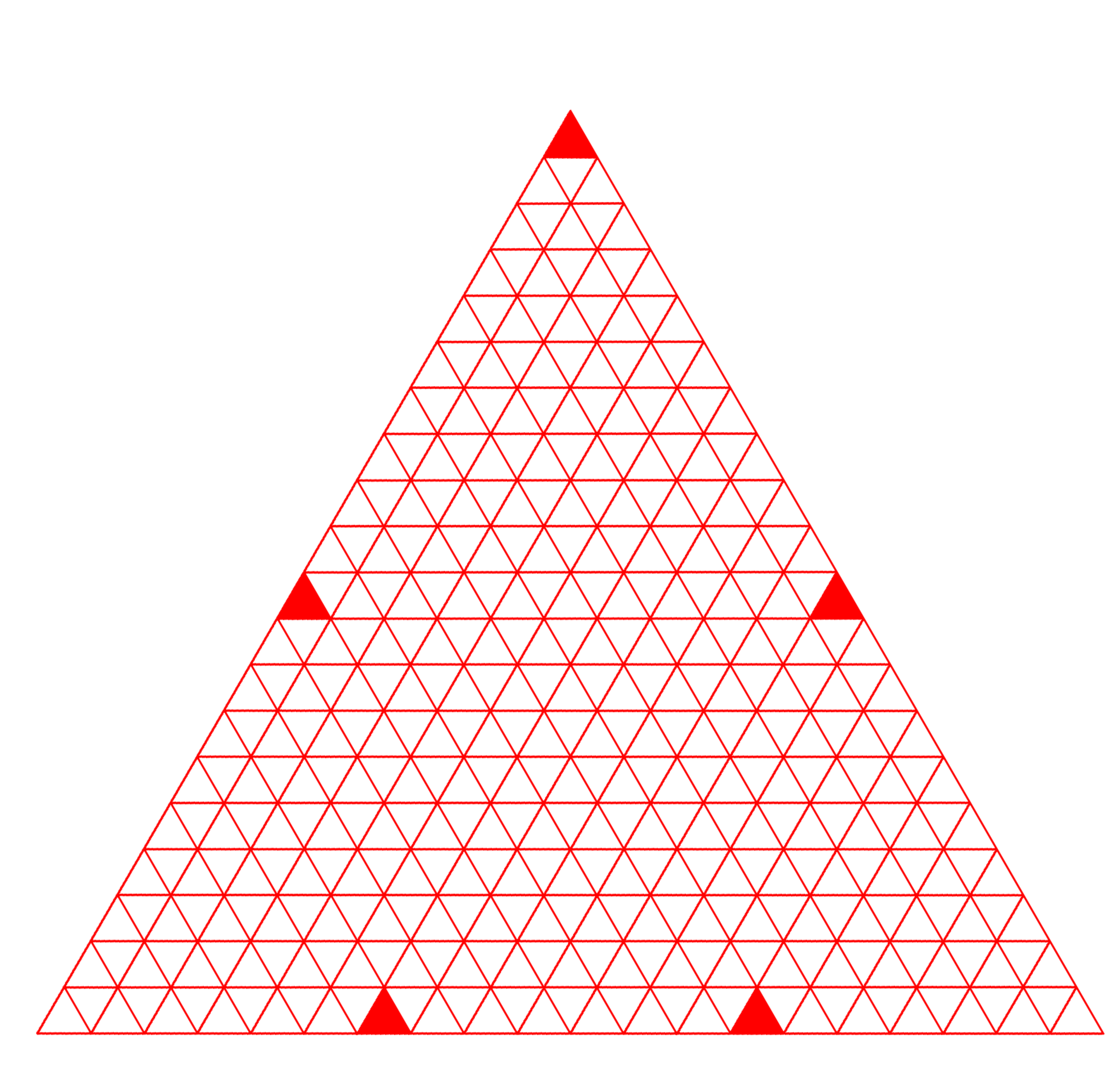}{triangles5}
\pgfdeclareimage[height=6.4cm,width=6.4cm]{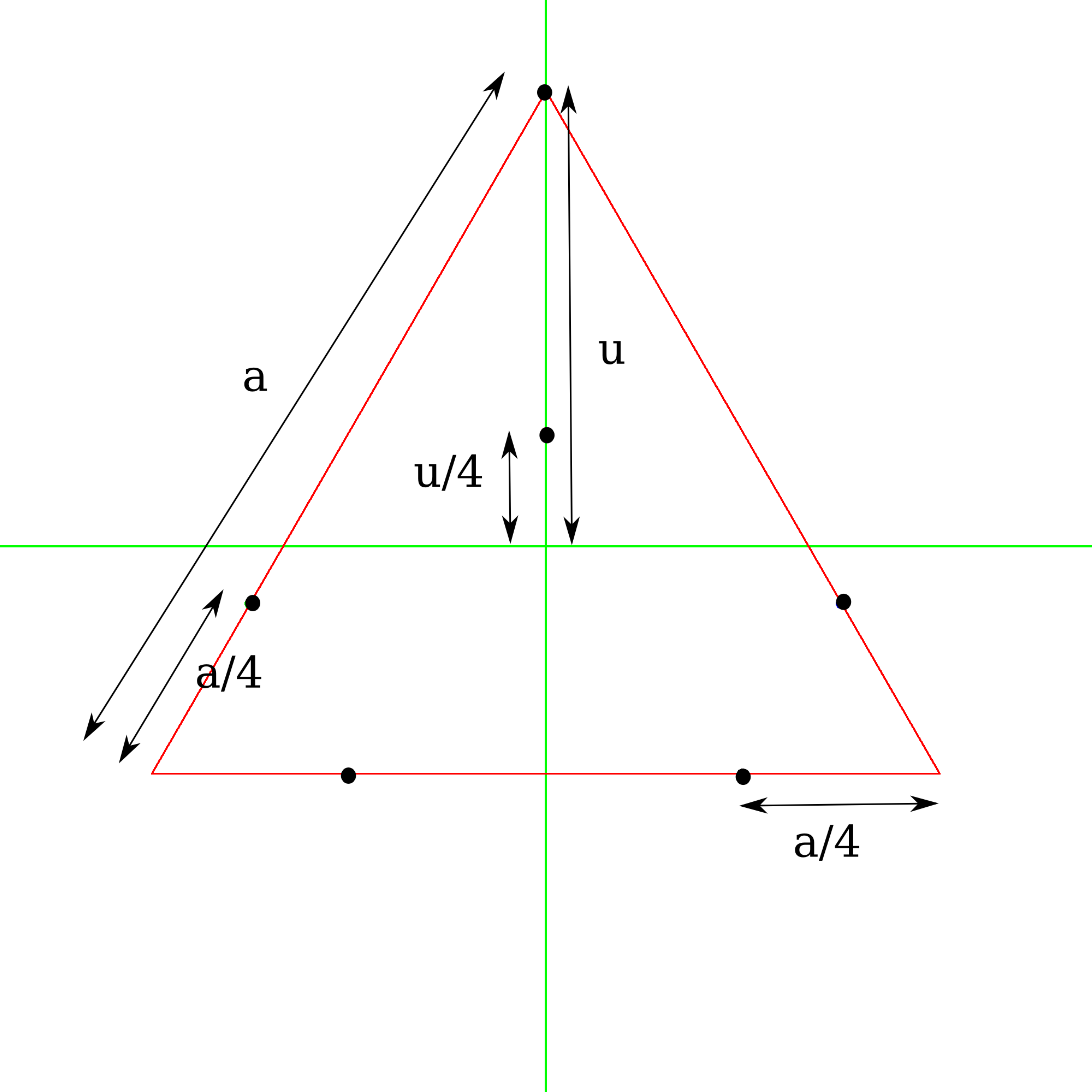}{sixpointsV1}
\pgfdeclareimage[height=6cm,width=6cm]{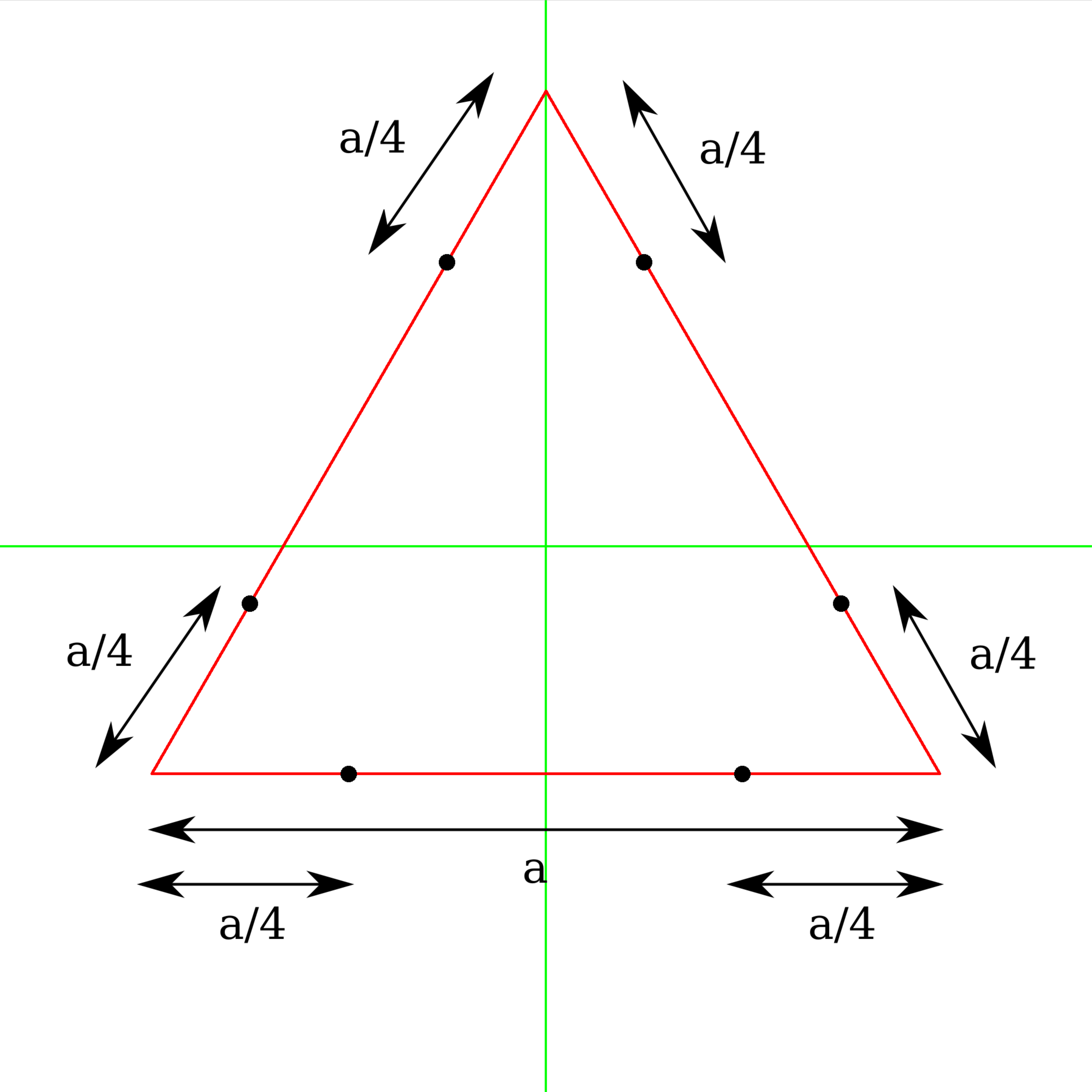}{sixpointsV2}
\pgfdeclareimage[height=6cm,width=6cm]{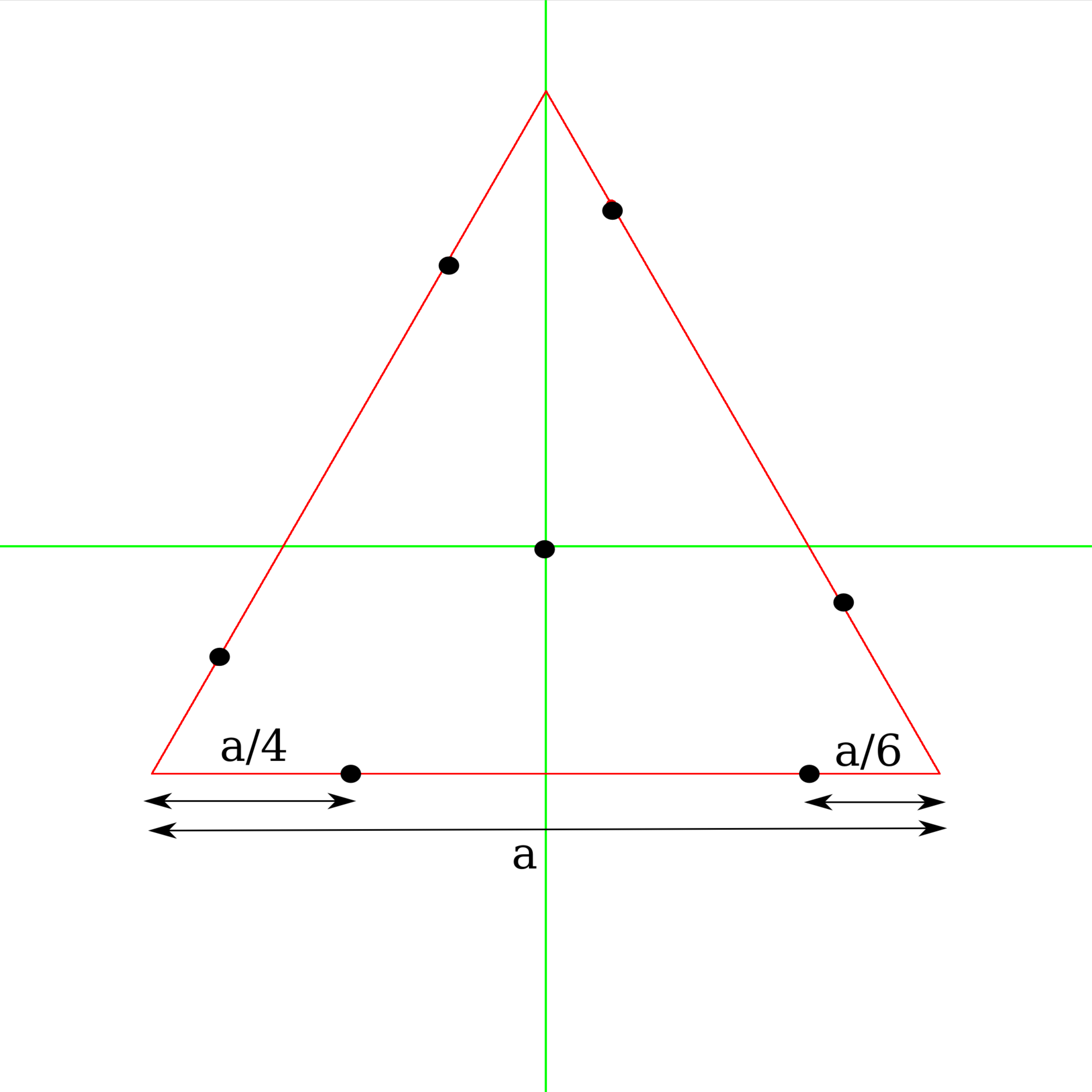}{conj7}

\title{Automated Proofs in Geometry : Computing Upper Bounds for the Heilbronn Problem for Triangles} 
\author{\small Francesco De Comit\'e    \\
\small Laboratoire d'Informatique Fondamentale de Lille \\
\small University of Science and Technology Lille (France)\\
\small{\email{Francesco.De-Comite@univ-lille1.fr}
}
 \and 
\small Jean-Paul Delahaye \\ 
\small Laboratoire d'Informatique Fondamentale de Lille \\
\small University of Science and Technology Lille (France)\\
\small{\email{delahaye@lifl.fr}}
}

\begin{document}

\fontfamily{phv}

\maketitle

\bibliographystyle{alpha}

\abstract{
\parindent 0pt
The Heilbronn problem for triangle\cite{Weisstein} is defined as follows: place $N$ points inside a triangle of unit area, so as to maximize the area of the smallest triangle obtained by choosing 3 points among $N$.
Several authors worked towards finding lower bounds or optimal configurations of points.

In this paper, we propose upper bounds for those problems, obtained by a method of automated theorem proving. 

}

\section{Introduction}\parindent 0pt
The Heilbronn problem for triangles can be expressed in the following way: 
\begin{quotation}
What is the minimum s of all $\sigma$ such that among every five points in a triangle of unit area, some three of them form a triangle of
area less than or equal to $\sigma$ ?
\end{quotation}

\begin{figure}[h]
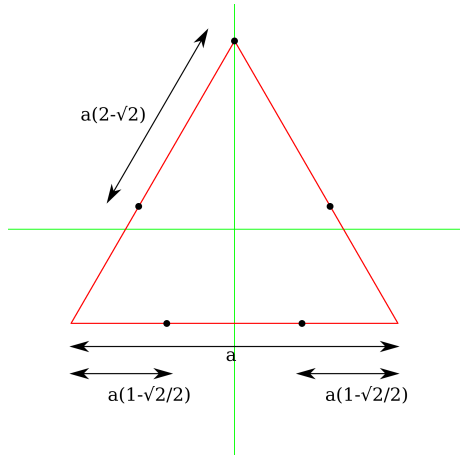

\begin{center}
\pgfuseimage{notreconjecture5points}
\caption{optimal solution for N=5: $s=3\sqrt{2}-2$\label{optimal5}}
\end{center}
\end{figure}

Optimal solutions for $N=3,4$ are trivial.  
Alexander Soifer pointed out that the proof of optimality for $N=5$ (figure \ref{optimal5})  was found by Royce Peng in 1989, and the outline of his proof is given in \cite{Soifer} (section 9.3, Problem 9.3.2).
Independantly, L.Yang, J.Z.Zhang and Z.B.Zeng\cite{Yang94} have found optimal solutions and proof of optimality for $N=5,6$ (figures \ref{sixpointsV1} and \ref{sixpointsV2}). David Cantrell\cite{Cantrell} has found lower bounds for $7\leq N\leq 16$, but no proof of optimality. 
Figure \ref{conj7} shows the conjectured optimal solution for $N=7$.
Matthew Kahle \cite{Kahle} derived upper bounds for $N=5$. We generalize his method, and computed new upper bounds for $N=5$ to $7$.

\begin{figure}
\begin{minipage}[b]{0.5\linewidth} 
\centering
\pgfuseimage{sixpointsV1}
\caption{Optimal solution for  $N=6: \sigma(6)={1\over{8}}$\label{sixpointsV1}}
\end{minipage}
\hspace{0.5cm} 
\begin{minipage}[b]{0.5\linewidth}
\centering
\pgfuseimage{sixpointsV2}
\caption{Optimal solution for $N=6$: $\sigma(6)={1\over{8}}$\label{sixpointsV2}}
\end{minipage}
\end{figure}

\begin{figure}[h]
\begin{center}
\pgfuseimage{conj7}
\caption{Conjecture for $N=7$: $\sigma(7)={7\over{72}}=0.0922\dots$\label{conj7}}
\end{center}
\end{figure}

\subsection{An automated proof of a new upper bound for $s$}
In order to obtain an upper bound for $s$, we  wrote a program inspired by  Kahle's proof\cite{Kahle}. The method is here described for $N=5$, but it is general and can be applied to any value of $N$. 

We divide the unit triangle T into P*P elementary triangles congruent to T, by drawing lines parallel to each side of T. 
We consider every 5-uples of triangles~:
\begin{itemize}
\item For each 5-uple, we consider every subset of three distinct triangles.
\item  For each subset of three triangles, we compute the maximum area of a triangle whose vertices are in each elementary triangle of the subset (this computation is 
made by an exhaustive enumeration of the 27 triangles defined by the vertices of the three elementary triangles).
\item The minimal area $\sigma(5)$ of such a triangle for this 5-uple is an upper bound of the $s$ associated to the family of 5-uples of points, when one point is taken in each of the five elementary triangles.
\item The maximal value of $\sigma(5)$ for all the 5-uples of triangles  we can define among P*P elementary triangles is an upper bound for $s$.
\end{itemize}

The enumeration is time-consuming, but elementary optimizations reduce the computing time.

As the result does not depend on the shape of the triangle, we can define T as an isoscele rectangle triangle
with small sides of length P: all the computations are then done with integers, results are also
integers. Hence $\sigma(5)$  is not subject to inferior rounding, and is a true upper bound for $s$. Our computation of $\sigma(5)$ as the upper bound for $s$ is then a kind of automated theorem proving method.

 We run experiments with increasing values of P, proving the  upper bound $$\sigma(5)=121/625=0.1936 \mbox{\  for \ } P=25$$

\begin{figure}[h]
\begin{center}
\begin{tabular}{|c|c|} \hline
P & upper bound\\ \hline
10 & $21/100=0.21$ \\ \hline
15& $46/225\approx 0.2044$ \\ \hline
20 & $79/400=0.1975$ \\ \hline
25 & $121/625=0.1936$\\ \hline
\end{tabular}
\caption{Evolution of the upper bound for $\sigma(5)$ as P increases, $N=5$\label{table1}.}
\end{center}
\end{figure}

For $P=10$, our upper bound of $0.21$ is slightly better than Kahle's result of $0.24$. The reason is that our method explored exhaustively each 5-uple of triangles, which is not the case for Kahle's reasoning. 

\begin{figure}[h]
\begin{center}
\pgfuseimage{triangles5}
\caption{An upper bound for $N=5: \sigma(5)\le {{121}\over{625}}$\label{triangles5}}
\end{center}
\end{figure}

\subsection{Improvement of the lower bound under some assumptions}

In order to improve the value of the lower bound, we re-ran the automated theorem proving program, for larger values of P, now assuming that the five {\em good} points
are on the edges and/or vertices of the initial triangle of unit area. 

This reduces the complexity of the computation, and leads to an upper bound of: $$87/500=0.174 \mbox{\ for\ } P=150$$

Thus, provided that it is true that the five points giving the value $s$ are on the perimeter of T, we can now write that we have a computer proof that: 

$$\sigma(5)\le 87/500$$ 

\begin{figure}[h]
\begin{minipage}[b]{0.5\linewidth} 
\centering
\rotatebox{180}{\includegraphics[width=7cm]{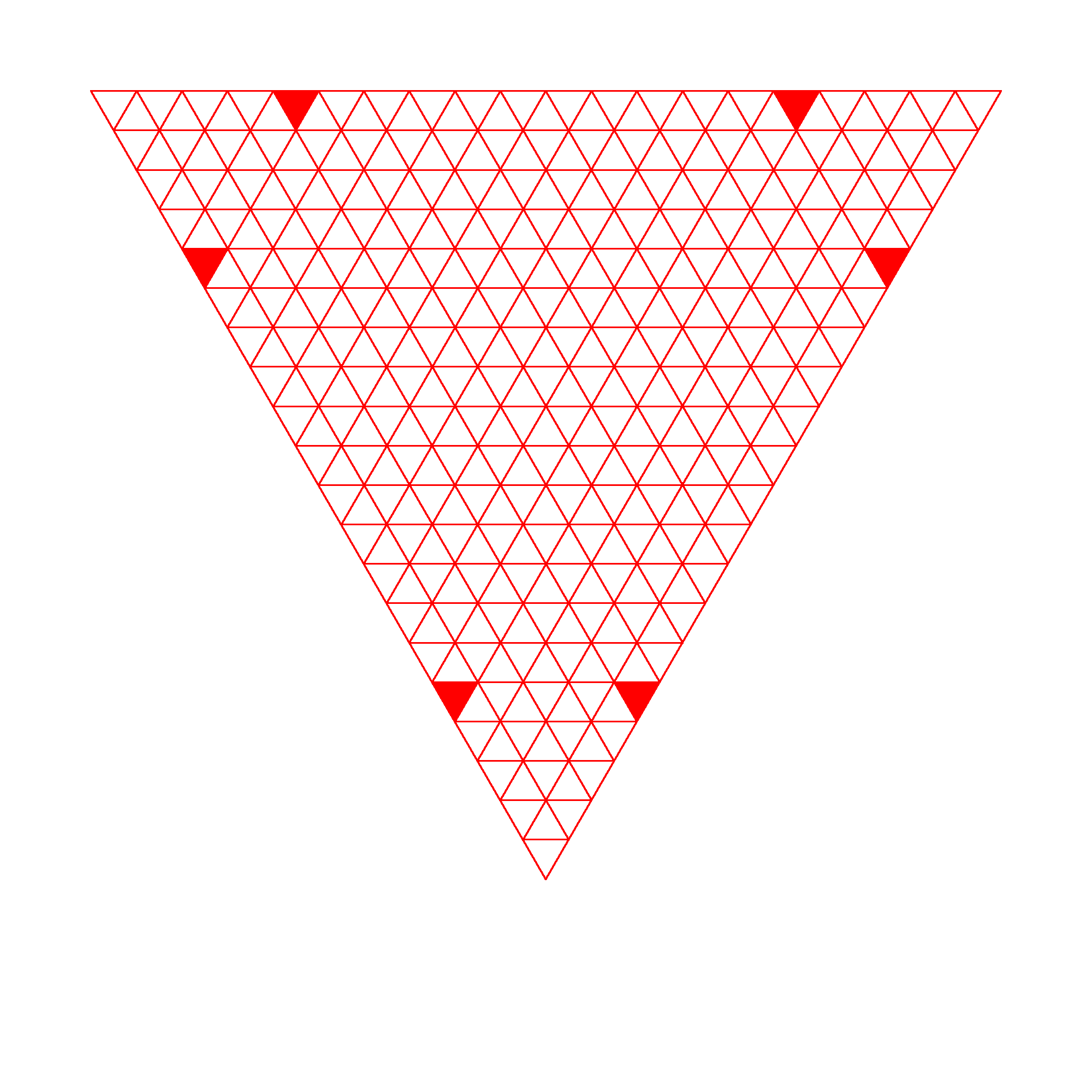}}
\caption{\label{maj6points}Upper bound for $N=6: \sigma(6)\le{3\over{20}}$}
\end{minipage}
\hspace{0.5cm} 
\begin{minipage}[b]{0.5\linewidth}
\centering
\rotatebox{180}{\includegraphics[width=7cm]{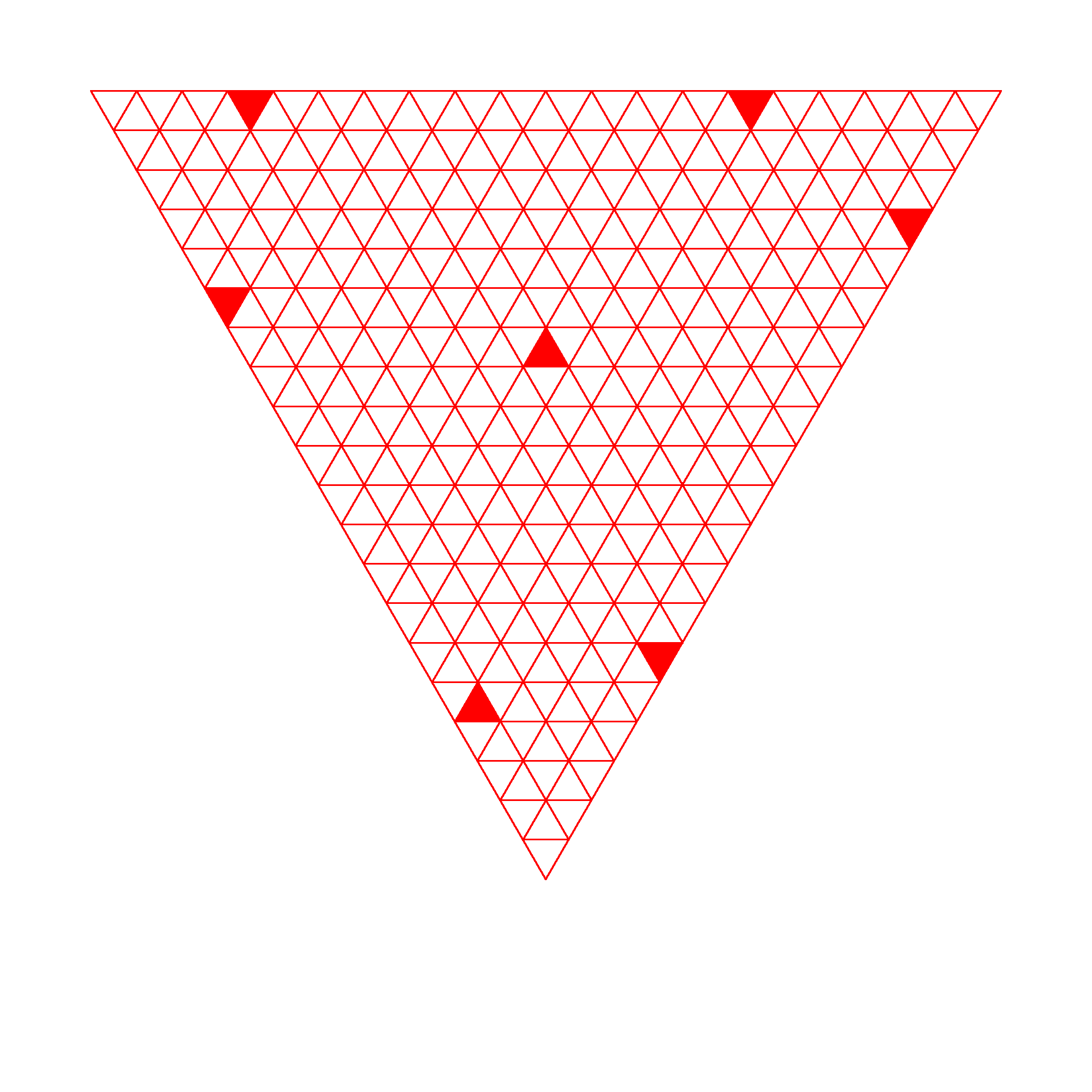}}
\caption{Upper bound for $N=7: \sigma(7)\le {{23}\over{200}}$\label{maj7points}}
\end{minipage}
\end{figure}

\section{Upper bounds on $s$ for $N=6$ and  $N=7$}

We investigated the cases $N=6$ and $N=7$, using the method previously used for $N=5$. Here are our results: 

\subsection*{N=6}

We proved the following bound: 
$$  \sigma(6)\le 3/20$$
Proof that $\sigma(6)\le 3/20$ is based on the configuration of Figure \ref{maj6points}. 

\subsection*{N=7}  $$\sigma(7)\le{23\over{200}}$$

Proof that $\sigma(7)\le 23/200$ is based on the configuration of Figure \ref{maj7points}.







\bibliography{biblio}

\end{document}